\documentclass[prl, twocolumn, showkeys, amsmath, amssymb]{revtex4}

\newcommand{\eexp}{\mbox{e}^}
\usepackage{graphicx}

\begin{document}

\title{The Effect of Mutators on Adaptability in Time-Varying Fitness Landscapes}

\author{Pavel Gorodetsky}
\email{pavelgor@bgu.ac.il}
\affiliation{Ben-Gurion University of the Negev, Be'er-Sheva, Israel}
\author{Emmanuel Tannenbaum}
\email{emanuelt@bgu.ac.il}
\affiliation{Ben-Gurion University of the Negev, Be'er-Sheva, Israel}

\begin{abstract}
This Letter studies the quasispecies dynamics of a population capable of genetic repair evolving on a time-dependent fitness landscape.  We develop a model that considers an asexual population of single-stranded, conservatively replicating genomes, whose only source of genetic variation is due to copying errors during replication.  We consider a time-dependent, single-fitness-peak landscape where the master sequence changes by a single point mutation every time $ \tau $.  We are able to analytically solve for the evolutionary dynamics of the population in the point-mutation limit.  In particular, our model provides an analytical expression for the fraction of mutators in the dynamic fitness landscape that agrees well with results from stochastic simulations.
\end{abstract}


\maketitle

Genetic repair plays an important role in maintaining the integrity of nearly all organismal genomes \cite{Voet}.  Two important examples of repair mechanisms are DNA proofreading and mismatch repair (MMR).  DNA proofreading is an error-correction mechanism that is built into the DNA replicases themselves:  If, during replication, the DNA replicase mistakenly pairs the wrong daughter base with the parent base, the replicase excises the daughter base and attaches a new daughter base.  DNA proofreading reduces the error rate from one mismatch every $ 10^4 $ bases to one mismatch every $ 10^7 $ bases \cite{Voet}.

Mismatch repair occurs immediately following daughter strand synthesis, and requires an ability to distinguish between the newly synthesized daughter strand and the parent strand.  In prokaryotes, this is achieved by exploiting the time lag between daughter strand synthesis and methylation of the daughter strand, while in eukaryotes parent-daughter strand discrimination is achieved via the presence of nicks in the daughter strand \cite{Voet}.

Mismatch repair reduces the error rate by a factor of $ 100 $, bringing the error rate in the wild-type down to one mismatch every $ 10^9 $ bases \cite{Voet}.  Mismatch repair is of particular interest to evolutionary biologists, because it is believed that mismatch-repair-deficient strains, or {\it mutators}, play a crucial role in the emergence of antibiotic drug resistance, and also act as gateway cells for the emergence of cancer \cite{Tannenbaum:03} (and references therein).  

In static environments, mutators are generally a relatively small fraction of the population \cite{LeClerc:96}, because their higher than wild-type mutation rate means that they accumulate deleterious mutations more quickly than non-mutators.  However, in dynamic environments, mutators can adapt more quickly than non-mutators, and so can rise to relatively large fractions of the population.  Experimentally, it is known that in static environments the mutator fraction of {\it Escherichia coli} is on the order of $ 0.1\% $, while pathogenic strains under pressure to respond to an adaptive immune system can have mutator fractions on the order of $ 10\% $ or more \cite{LeClerc:96}.

Because of their importance, mutators have been the subject of considerable experimental and theoretical work \cite{Tannenbaum:03, Tannenbaum:04, Nowak:03, Kessler:98, Travis:02, Painter:75,
Lenski:97, Palmer:06}.  Theoretical work on the subject has consisted both of numerical studies modeling the evolution of mutation rate in time-varying environments, as well as analytical work modeling how mutators influence adaptation to a new environment.  

In this Letter, we develop an analytically solvable model describing the influence of mutators on adaptation in time-varying environments.  Such a model is an extension of previous analytical work on the subject, which either focused on adaptation to a new, but static, environment \cite{Painter:75, Kessler:98}, or on adaptation in environments that alternate between two states \cite{Palmer:06}.  This Letter is also an extension of previous numerical work, in that we provide approximate analytical expressions for the fraction of mutators in dynamic environments.  The work here therefore provides a starting point for understanding the role of mutators in environments where co-evolutionary dynamics is important, such as is the case with an adaptive immune system.

An analytically solvable model describing quasispecies evolution in a dynamic fitness landscape was first developed in \cite{Nilsson:00}.  Following the approach in \cite{Nilsson:00}, we assume a single-peak fitness landscape in which there is one high fitness sequence, the master sequence $ \sigma_{via, 0} $, in an otherwise flat landscape.  We also assume that there exists a sequence $ \sigma_{rep,0} $ which corresponds to a working MMR mechanism.  Our organisms then have genomes that may be denoted by $ \sigma = \sigma_{via} \sigma_{rep} $, where $ \sigma_{via} $ is the genome region that controls viability and $ \sigma_{rep} $ is a region that controls repair.  $ \sigma_{via} $ consists of $ L_{via} $ bases, and $ \sigma_{rep} $ consists of $ L_{rep} $ bases,  so that the total genome length is $ L = L_{via} + L_{rep} $.  Note that we are essentially considering a two-locus model, which is the approach taken by previous authors \cite{Kessler:98, Painter:75}.  The difference here is that we are assuming master-sequence-based fitness and ``repair'' landscapes, so that, in contrast to previous work, various transition probabilities do not need to be considered as independent variables, but rather may be computed from per-base error probabilities and sequence lengths.

A viable organism, i.e. an organism for which $ \sigma_{via} = \sigma_{via, 0} $, has a first-order growth rate constant $ \kappa_{\sigma} = k > 1 $, while for an unviable organism, $ \kappa_{\sigma} = 1 $.  We define a repair landscape in an analogous manner:  $ \epsilon_{\sigma} = \epsilon_r \epsilon $ for organisms with $ \sigma_{rep} = \sigma_{rep, 0} $ and $ \epsilon_{\sigma} = \epsilon $ otherwise, where $ \epsilon $ is the per-base replication error probability and $ \epsilon_{r} $ is the per-base repair error probability.

To create a dynamic landscape, we move the fitness peak in the genotype space to one of its nearest neighbors (chosen randomly) at regular time intervals $ \tau $.  We then define the parameters $ n_{00} $, $ n_{01} $, $ n_{10} $ and $ n_{11} $ as follows: $ n_{00} $ is the number of viable non-mutators, $ n_{01} $ is the number of viable mutators, $ n_{10} $ is the number of non-mutators that will be viable after the next peak shift, and $ n_{11} $ is the number of mutators that will be viable after the next peak shift.

We assume that $ \epsilon $ is sufficiently small that only point mutations are important, and we also assume that $ n_{10} $ and $ n_{11} $ are zero immediately after a peak shift.  That is, when the fitness peak shifts to a new master sequence, the number of organisms that will be viable after the next peak shift is assumed to be negligible.

Based on these assumptions, we have that, during the time interval $ [n \tau, (n + 1) \tau) $ immediately following the $ n^{\mbox{th}} $ peak shift, the quasispecies equations take the form,
\begin{equation}
\frac{d}{dt}
\left(\begin{array}{c}
n_{00} \\
n_{01} \\
n_{10} \\
n_{11}
\end{array}
\right)
 = 
\left(\begin{array}{cccc}
C_{1} & 0 & 0 & 0\\ 
C_{3} & C_{2} & 0 & 0\\
C_{5} & 0 & C_{4} & 0\\
0 & C_7 & C_8 & C_6
\end{array}
\right)
\left(\begin{array}{c}
n_{00} \\
n_{01} \\
n_{10} \\
n_{11}
\end{array}
\right)
\end{equation}
where the $ C_i $ coefficients are defined as,
\begin{eqnarray}
&   &
C_{1} = k\left(1-\epsilon\epsilon_{r}\right)^{L} 
\nonumber \\
&   &
C_{2} = k\left(1-\epsilon\right)^{L_{via}}
\nonumber \\
&   &
C_{3} = k\left(1-\epsilon\epsilon_{r}\right)^{L_{via}}[1-\left(1-\epsilon\epsilon_{r}\right)^{L_{rep}}]
\nonumber \\
&   &
C_{4} = \left(1-\epsilon\epsilon_{r}\right)^{L}
\nonumber \\
&   &
C_{5} = \frac{k\epsilon\epsilon_{r}}{S-1}\left(1-\epsilon\epsilon_{r}\right)^{L-1}
\nonumber \\
&   &
C_{6} = (1-\epsilon)^{L_{via}}
\nonumber \\
&   &
C_{7} = \frac{k\epsilon}{S-1}(1-\epsilon)^{L_{via}}
\nonumber \\
&   &
C_{8} = (1-\epsilon\epsilon_{r})^{L_{via}}[1-(1-\epsilon\epsilon_{r})^{L_{rep}}]
\end{eqnarray}

We may solve Eq. (1) for $ t \in [n \tau, (n+1)t) $ with the initial conditions $ n_{10}(n \tau) = n_{11}(n \tau) = 0 $.  Noting that at $ t = (n + 1) \tau $ the fitness peak shifts to the next master sequence, we have,
\begin{eqnarray}
&   &
n_{00}((n + 1) \tau) = A_{10}(\tau) n_{00}(n \tau) 
\nonumber \\
&   &
n_{01}((n + 1) \tau) = A_{11}(\tau) n_{00}(n \tau) + B_{11}(\tau) n_{01}(n \tau)
\nonumber \\
\end{eqnarray}
where the coefficients $ A_{10}(\tau) $, $ A_{11}(\tau) $, $ B_{11}(\tau) $ are given by,
\begin{eqnarray}
&   &
A_{10}(\tau) = \frac{\left( \eexp{k \tau C_4} - \eexp{\tau C_4}\right)C_5}{\left(k-1\right)C_4}
\nonumber \\
&   &
A_{11}(\tau) =  
\frac{(\eexp{\tau C_4} - \eexp{k\tau C_4})C_5 C_6 C_8}{(k-1)(C_4 - C_6)(kC_4 - C_6) C_4} 
\nonumber \\
&   &
+ 
\frac{k(\eexp{\tau C_6} - \eexp{k\tau C_6})C_4 C_7 C_8}{(k-1)(C_4 - C_6)(kC_4 - C_6) C_6} 
\nonumber \\
&   &
+ \frac{[\eexp{k\tau C_4} - \eexp{\tau C_6} + k(\eexp{\tau C_6} - \eexp{\tau C_4})] C_5 C_8 }
{(k-1)(C_4 - C_6)(kC_4 - C_6)} 
\nonumber \\
&   &
+ \frac{[(k-1)\eexp{k\tau C_4} - k\eexp{\tau C_6} + \eexp{k\tau C_6}]C_7 C_8}{(k-1)(C_4 - C_6)(kC_4 - C_6)C_4C_6}
\nonumber \\
&   &
B_{11}(\tau) = \frac{\left( \eexp{k\tau C_6} - \eexp{\tau C_6} \right) C_7}{\left(k - 1 \right)C_6}
\end{eqnarray}

Let us denote by $ r_n $ the ratio of viable mutators to viable non-mutators immediately following the $ n^{\mbox{th}} $ peak shift.  Then from the solution given above we have,
\begin{equation}
r_{n + 1} = \alpha(\tau) + \beta(\tau) r_n 
\end{equation}
where $ \alpha(\tau) \equiv A_{11}(\tau)/A_{10}(\tau) $ and $ \beta(\tau) = B_{11}(\tau)/A_{10}(\tau) $, from which it can be shown that $ r_n = \alpha(\tau) \sum_{i = 0}^{n-1} [\beta(\tau)]^{i} + \beta^n(\tau) r_0 $. If $ 0 \leq \beta < 1 $, then, as $ n \rightarrow \infty $, we obtain that $ r_{\infty} \equiv \lim_{n \rightarrow \infty} r_n = \frac{A_{11}(\tau)}{A_{10}(\tau) - B_{11}(\tau)} $, which implies a periodic solution where the fraction of viable organisms that are non-mutators at the beginning of each cycle is nonzero.  When $ \beta \geq 1 $ we obtain that $ r_n $ diverges, and so the fraction of viable organisms that are mutators goes to $ 1 $.  This phenomenon is known as the {\it repair catastrophe} \cite{Tannenbaum:03}.  

For the periodic solution that develops after a sufficient number of iterations, the fraction of viable organisms that are mutators at the beginning of a peak shift is then $ r_{\infty}/(1 + r_{\infty}) $.  As $ \tau \rightarrow \infty $, it may be readily shown that the fraction of viable organisms that are mutators just before the next peak shift is given by the static landscape expression obtained in \cite{Tannenbaum:04}. 

In order for the population to form a stable quasispecies and remain viable in the time-varying landscape, the relative growth of the viable population between the peak shifts should be larger than that of the background (genome sequences some distance away from the master sequence).  Since the growth constant of non-viable organisms is equal to $ 1 $, the criterion for adaptability is given by,
\begin{equation}
\frac{n_{00}((n + 1) \tau) + n_{01}((n + 1) \tau)}{n_{00}(n \tau) + n_{01}(n \tau)} > \eexp{\tau}
\label{CriterionForAdaptability}
\end{equation}
If this condition is not met there will be a drift from the master sequence to the ``background'', corresponding to the error catastrophe.

When the fraction of viable organisms that are non-mutators is nonzero, Eq. (6) may be shown to be equivalent to,
\begin{equation}
\kappa_{nm} \equiv \frac{k}{k-1}\frac{\epsilon\epsilon_{r}}{1-\epsilon\epsilon_{r}}\frac{\eexp{[k(1-\epsilon\epsilon_{r})^{L}-1]\tau} - \eexp{[(1-\epsilon\epsilon_{r})^{L}-1]\tau}}{S-1}
\end{equation}
while when the fraction of viable organisms that are non-mutators is zero, i.e. when the population has undergone the repair catastrophe, Eq. (6) may be shown to be equivalent to,
\begin{equation}
\kappa_{m} \equiv \frac{k}{k-1}\frac{\epsilon}{1-\epsilon}\frac{\eexp{[k(1-\epsilon)^{L_{via}}-1]\tau}
-\eexp{[(1-\epsilon)^{L_{via}}-1]\tau}}{S-1}
\end{equation}

We may define three distinct parameter regimes, each corresponding to the viability or non-viability of the non-mutator and mutator populations, respectively.

The first region is defined by $ \kappa_{nm} > 1, \kappa_m $.  Here, the non-mutators' effective growth rate is both larger than that of the background as well as that of the mutators, so that a stable quasispecies is formed with a non-zero fraction of viable non-mutators.

The second region is defined by $ \kappa_{m} > 1, \kappa_{nm} $.  Here the mutators' effective growth is both larger than that of the background as well as that of the non-mutators, so that a stable quasispecies is formed, but it consists entirely of mutators.  The transition between the first and second regions corresponds to a localization to delocalization transition over the repair portion of the genome that is termed the repair catastrophe \cite{Tannenbaum:03}.

Finally, the third region is defined by $ \kappa_{nm}, \kappa_m < 1 $, so that no stable quasispecies forms, and the population is unable to adapt to the changing fitness landscape.

The solutions to the equations $ \kappa_{nm}, \kappa_{m} = 1 $ set the upper and lower mutation thresholds for the population to survive (see Figure 1).  The upper mutation threshold is the ordinary error catastrophe for the static landscape, while the lower threshold appears only in the case of a dynamic landscape. This lower threshold arises because, in a dynamic fitness landscape, the mutation rate must have a minimal value to allow the population to adapt.  Below this dynamic error threshold, the population is unable to adapt, and so no stable quasispecies is formed \cite{Nilsson:00}.

There is some minimal value of $ \tau $, denoted $ \tau_{min} $, below which the population cannot
adapt.  This corresponds to a threshold rate of change of the fitness landscape above which the population cannot adapt, irrespective of the error threshold.  This minimal value for $ \tau $ arises because, as $ \tau $ is decreased, the mutation rate must increase to maintain a stable quasispecies.  Once this minimal mutation rate exceeds the static error threshold, then no quasispecies can exist.

An analytical approximation for $ \tau_{min} $ can be found by expanding Eq. (8) to second-order in $ \epsilon $, solving for $ \epsilon $, and then solving for the value of $ \tau $ where the two solutions are equal.  Assuming that $ k $ is sufficiently larger than $ 1 $, and that $ \tau $ is sufficiently large, so that $ \eexp{k \tau} >> \eexp{\tau} $, we obtain the following approximate expression for $ \tau_{min} $:
\begin{equation}
\tau_{min} \approx \frac{1}{k}\text{ln}\left[4(k-1)(S-1)L_{via}\right]
\label{tauMinEq}
\end{equation}

An $ \epsilon $ versus $ \tau $ plot of the solution to the equation $ \kappa_{nm} = \kappa_m $ yields a continuous curve, above which it can be shown that $ \kappa_{nm} > \kappa_m $, and below which it can be shown that $ \kappa_{nm} < \kappa_m $.  Therefore, the region above this curve that also lies between the lower and upper mutational thresholds of the non-mutators corresponds to a parameter regime where there is a stable quasispecies with a nonzero fraction of non-mutators.  The region below this curve that also lies between the lower and upper mutational thresholds of the mutators corresponds to a parameter regime where there is a stable quasispecies consisting entirely of mutators.

\begin{figure}
\includegraphics[width = 0.8\linewidth, angle = 90]{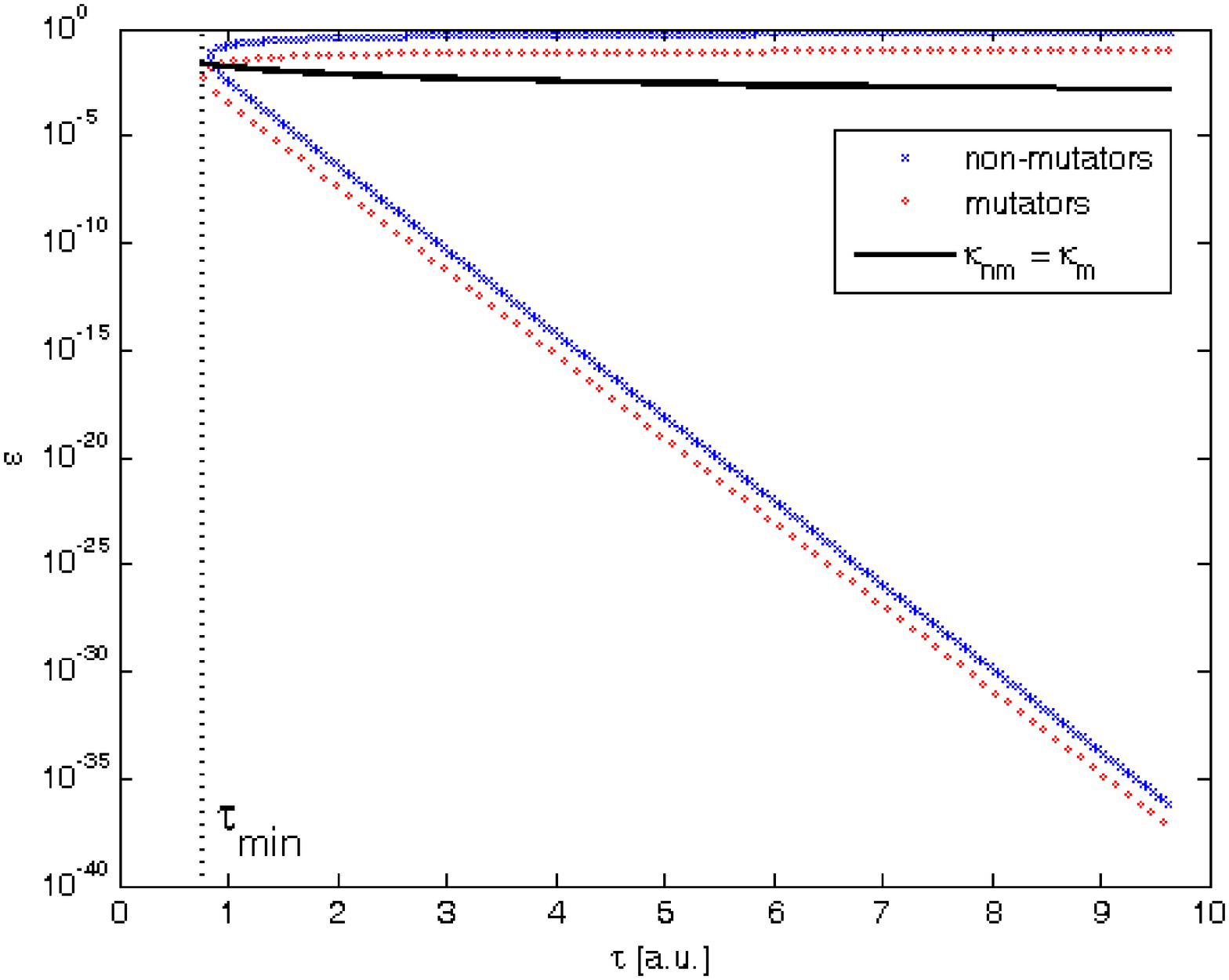}
\caption{$ L_{via} = 20 $, $ L_{rep} = 10 $, $ S = 4 $, $ \epsilon_r = 0.1 $:  Solution of $ \kappa_{nm} = 1 $ and $ \kappa_{m} = 1 $ as a function of cycle length $ \tau $.  The area confined by the symbols is the survivability region of the population.  In the region above the solid curve the fraction of non-mutators is nonzero, while below the solid curve only mutators exist.  $ \tau_{min} $ found from Eq. (9).}
\end{figure}

\begin{figure}
\includegraphics[width = 0.8\linewidth, angle = 90]{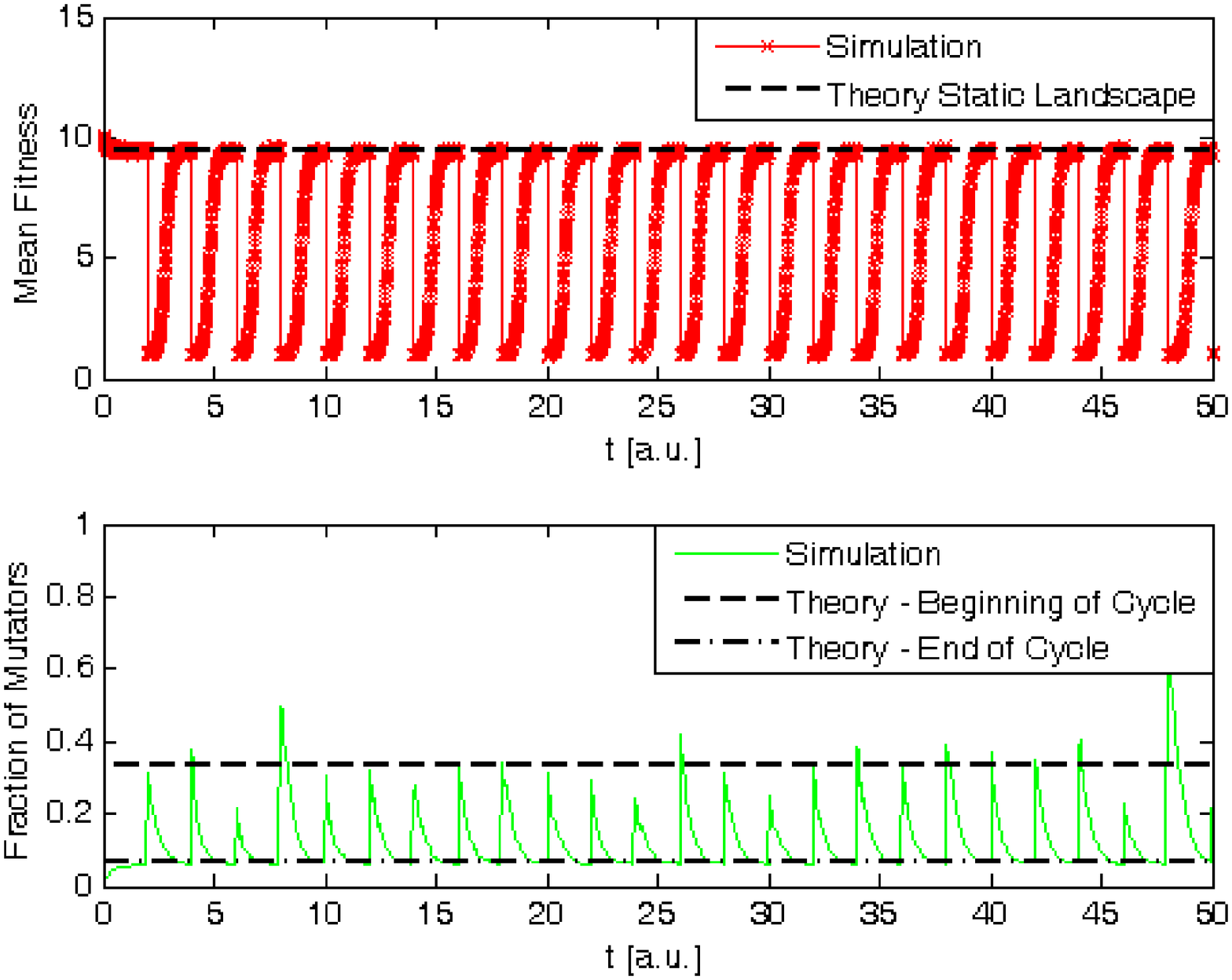}
\caption{$ N = 1.5 \times 10^5 $, $ L_{via} = 20 $, $ L_{rep} = 10 $, $ S = 4 $, $ \epsilon = 0.02 $, 
$ \epsilon_r = 0.1 $, $ \tau = 2 $:  A single trajectory from a stochastic simulation.  Note that the mean fitness reaches its static landscape value.  The simulated fraction of mutators at the beginning and at the end of each cycle shows good agreement with the analytical results.}
\end{figure}

\begin{figure}
\includegraphics[width = 0.8\linewidth, angle = 90]{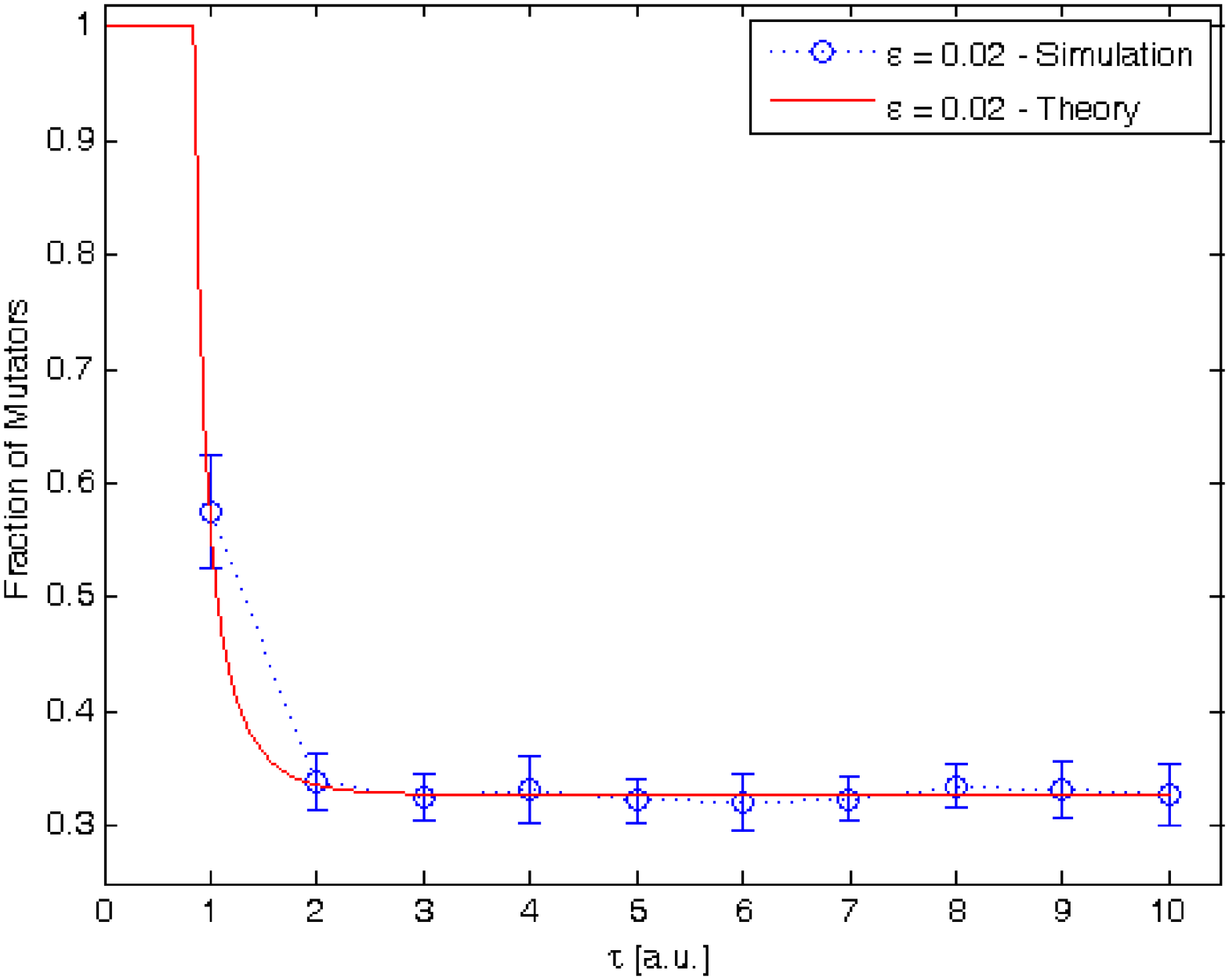}
\caption{$ N = 1.5 \times 10^5 $, $ L_{via} = 20 $, $ L_{rep} = 10 $, $ S = 4 $, $ \epsilon = 0.02 $, 
$ \epsilon_{r} = 0.1 $:  Fraction of mutators as a function of $ \tau $.  The results are averaged over 
$ 10 $ independent runs.}
\end{figure}

We developed a stochastic code to simulate the dynamics of a population of self-replicating genomes of the form $ \sigma = \sigma_{via} \sigma_{rep} $.  At each time step, a given genome replicates with probability $ \kappa_{\sigma} \Delta t $, where $ \kappa_{\sigma} $ is the growth constant and 
$ \Delta t $ is the size of the time step.  We choose $ \Delta t $ to be sufficiently small so that the replication probability is much smaller than one.  After every time interval $ \tau $, a shift occurs and a new master sequence is generated to be one point mutation away from the old one.  The mutation is generated at a different random location in the viable region to avoid population accumulation (to meet the initial conditions of the analytical model), i.e. the mutation position repeats itself every $ L_{via} $ peak shifts. Each realization simulated at least $ 10^{4} $ time steps to make sure a periodic solution was reached. 

As can be seen from Figures 2 and 3, as long as $ \epsilon $ is sufficiently small, we obtain good agreement between theory and simulation.  However, due to stochastic effects, the fraction of mutators requires a larger population size than other evolution parameters (e.g. mean fitness, mean Hamming distance) before reasonable agreement with the theoretical results is obtained.

For sufficiently large $ \epsilon $, the agreement between our point-mutation model and the simulation results breaks down.  The reason for this is that, when $ \epsilon $ is small, the $ n_{00} \rightarrow n_{01} $, $ n_{00} \rightarrow n_{10} $, and $ n_{01} \rightarrow n_{11} $ transition probabilities are all first-order in $ \epsilon $, while the transition probability $ n_{00} \rightarrow n_{11} $ is second-order in $ \epsilon $, and is therefore neglected in a model that only considers mutation probabilities that are up to first-order in $ \epsilon $.

For larger $ \epsilon $, the non-mutator to mutator transition probability, which is $ 1 - (1 - \epsilon_r \epsilon)^{L_{rep}} $, goes from being first-order in $ \epsilon $ to zeroth-order, so that the $ n_{00} \rightarrow n_{11} $ becomes first-order in $ \epsilon $ as well.  Therefore, if we wish to use a first-order model for larger $ \epsilon $, then this additonal transition probability must be included.  Including the $ n_{00} \rightarrow n_{11} $ transition for larger $ \epsilon $ significantly improves the theoretical prediction of the mutator fraction.  For future research, we wish to develop a more systematic approach for estimating the mutator fraction in a dynamic fitness landscape that does not require the assumption of small $ \epsilon $.  Furthermore, we would like to move beyond mean-field descriptions of the evolutionary dynamics and develop analytical approaches for quantifying various stochastic effects.

This research was supported by the Israel Science Foundation (Alon Fellowship) and by the United States - Israel Binational Science Foundation.

\end{document}